\documentclass[conference]{IEEEtran}
\IEEEoverridecommandlockouts
\usepackage{cite}
\usepackage{amsmath,amssymb,amsfonts}
\usepackage{algorithmic}
\usepackage{graphicx}
\usepackage{textcomp}
\usepackage{xcolor}
\usepackage{tabularx}
\usepackage{booktabs}

\usepackage{tikz}
\usetikzlibrary{arrows.meta,positioning}

\definecolor{ecpblue}{HTML}{1F4E79}
\definecolor{ecpbluebg}{HTML}{E7EFF7}
\definecolor{ecpgreen}{HTML}{2E6F53}
\definecolor{ecpgreenbg}{HTML}{E4F0EA}
\definecolor{ecpamber}{HTML}{9C5A16}
\definecolor{ecpamberbg}{HTML}{FAEEE1}
\definecolor{ecpplum}{HTML}{6B3F72}
\definecolor{ecpplumbg}{HTML}{F1E9F3}
\definecolor{ecpgray}{HTML}{4A4A4A}
\definecolor{ecpgraybg}{HTML}{F1F1F1}

\tikzset{
  ecpbox/.style    = {draw=ecpgray, rounded corners=1.6pt, align=center,
                      inner xsep=3pt, inner ysep=3pt, fill=white},
  boxblue/.style   = {ecpbox, draw=ecpblue,  fill=ecpbluebg},
  boxgreen/.style  = {ecpbox, draw=ecpgreen, fill=ecpgreenbg},
  boxamber/.style  = {ecpbox, draw=ecpamber, fill=ecpamberbg},
  boxplum/.style   = {ecpbox, draw=ecpplum,  fill=ecpplumbg},
  boxgray/.style   = {ecpbox, draw=ecpgray,  fill=ecpgraybg},
  ecparrow/.style  = {-{Latex[length=2mm,width=1.6mm]}, ecpgray, semithick},
  ecpdash/.style   = {ecparrow, dashed},
  ecplabel/.style  = {align=center, inner sep=1.5pt, font=\scriptsize},
}
\def\BibTeX{{\rm B\kern-.05em{\sc i\kern-.025em b}\kern-.08em
    T\kern-.1667em\lower.7ex\hbox{E}\kern-.125emX}}

\begin{document}

\title{The Evaluation Context Protocol (ECP): A Portable Contract for AI Agent Evaluation  }

\author{
\IEEEauthorblockN{Aniket Wattamwar}
\IEEEauthorblockA{\textit{aniket.wattamwar17@gmail.com} \\
}
\and
\IEEEauthorblockN{Manav Anandani}
\IEEEauthorblockA{\textit{manavanandani304@gmail.com} \\
}
\and 
\IEEEauthorblockN{Mrunal Kakirwar}
\IEEEauthorblockA{\textit{kakirwarm@gmail.com} \\
}
}

\maketitle

\begin{abstract}
The evolution of artificial intelligence has necessitated a fundamental shift from evaluating isolated Large Language Models (LLMs) to assessing autonomous agentic architectures. This paper explores the critical methodologies for evaluating AI agents and the essential role of advanced observability infrastructure. We analyze the architectural components of agents and identify the severe limitations of current evaluation paradigms, including benchmark exploitation, the "confidently wrong" phenomenon, and the discrepancy between theoretical capability and operational reliability. To begin addressing the fragmentation in current evaluation infrastructure, this paper proposes the Evaluation Context Protocol (ECP), an early-stage, vendor-neutral framework intended to act as a portable evaluation contract layer for agentic systems. In its current form ECP defines a small JSON-RPC interface over which an agent exposes its user-visible output, the tool calls it made, and evaluator-safe audit context, and against which programmatic checks can be run uniformly across frameworks and continuous integration systems. We describe an open-source reference implementation that includes adapters for LangChain, LlamaIndex, CrewAI, and PydanticAI, and we situate the design against failure modes documented in the recent literature. ECP is presented as work in progress rather than a finished standard: the evaluation surface, method set, and grader families are all expected to change as the protocol is exercised against more systems, and the empirical validation required to justify adoption is outlined as future work.
\end{abstract}


\section{Introduction to the Agentic Evaluation Paradigm}

The evolution of artificial intelligence over the past several years has necessitated a fundamental architectural and philosophical shift in system design. Historically, the field has focused on isolated, stateless Large Language Models (LLMs) that function primarily as static response generators. In these traditional setups, a user provides a prompt, and the model returns a single, probabilistic text output based on its pre-trained weights. However, the industry has rapidly transitioned toward autonomous agentic architectures \cite{b31}. An AI agent is a dynamic behavioral system that integrates a foundation model with advanced reasoning capabilities, explicit planning modules, persistent memory mechanisms, and the ability to autonomously execute external tools to interact with its environment \cite{b23}. 

For researchers and developers new to this domain, it is critical to understand that evaluating an AI agent is a vastly different scientific problem than evaluating a standard language model. A language model can often be assessed by scoring a single response against a predefined benchmark, a reference answer, or a preference judgment \cite{b22}. If a standard chatbot hallucinates an answer, the error is confined to the text; the user can simply ignore it or issue a correction \cite{b3,b15}. An AI agent, however, takes actions with tangible consequences. It may execute SQL queries, alter database states, manage vendor communications, or navigate web interfaces \cite{b10,b12}. A hallucinated tool invocation or an incorrect reasoning step within an agentic pipeline can corrupt enterprise data, trigger unauthorized financial transactions, or compromise the security of entire digital systems \cite{b10}.

Given their inherent autonomy, evaluating these systems is critical to ensuring they function correctly, adhere to ethical principles, and resist manipulation \cite{b6}. The capabilities that make agents highly autonomous and useful in the real world such as multi-step reasoning, uncertainty handling, and dynamic contextual adaptation are precisely the characteristics that make them exceptionally difficult to evaluate reliably \cite{b8,b27}. The evaluation must capture the ``trajectory,'' which represents the end-to-end sequence of internal reasoning, tool calls, and environmental observations \cite{b8}.

This comprehensive report explores the current methodologies for evaluating AI agents and the critical necessity of advanced observability infrastructure. It exhaustively details where and why current evaluation paradigms fail. To begin addressing the fragmentation in agent evaluation contracts, this paper culminates in proposing the Evaluation Context Protocol (ECP), an early-stage, vendor-neutral framework intended as a portable evaluation contract layer for agentic systems. We emphasize at the outset that ECP is an initial proposal accompanied by a working reference implementation, not a settled standard, and that the design presented here is expected to evolve substantially.

\section{Fundamentals of Agent Architecture and Observability}

To establish a rigorous evaluation methodology, one must first deconstruct the architecture of an AI agent and define what must be observed. At the system level, an agent is modularized into distinct components with clear operational contracts.

The core of the agent is typically an LLM policy module, which serves as the central reasoning engine \cite{b2,b14,b15}. Surrounding this core are specialized modules: planners that decompose complex user intents into sequential steps, tool routers that determine which external APIs to invoke, and memory modules that preserve context across long horizons \cite{b9}. Memory is subdivided into episodic memory (what happened previously), semantic memory (factual knowledge), and procedural memory (skills and operational instructions), all of which support coherence beyond the constraints of a raw context window \cite{b10}. Furthermore, advanced architectures employ critic or verifier modules to cross-check claims against tool outputs and trusted sources, mitigating the risk of compounding errors \cite{b10}.

Evaluating this interconnected architecture requires deep observability. Observability in this context is the practice of instrumenting the agent to emit structured telemetry data logs, traces, and spans that make its internal state and decision-making processes transparent to human operators and automated grading systems. True agent observability demands tracking complete trajectories. A complete trajectory includes the agent's high-level plans and subgoals, every single tool invocation with its corresponding parameters and environmental responses, the intermediate reasoning steps justifying each action, and the final output alongside any environmental side effects, such as database writes or file modifications \cite{b7}.

Without this level of observability, an evaluator is effectively blind. If an agent arrives at the correct final answer but does so by utilizing a highly inefficient path, by hallucinating intermediate data that coincidentally aligns with the truth, or by inappropriately accessing restricted tools, the system is fundamentally flawed despite the ``correct'' outcome \cite{b4}. Observability transforms the ``black box'' of agent execution into a transparent sequence of auditable events.

\section{Current Evaluation Methods for AI Agents}

The current landscape of AI agent evaluation encompasses a broad spectrum of methodologies, ranging from static capability testing to dynamic simulations and automated grading systems. These methods attempt to measure various outcomes, including task completion, interaction quality, and system efficiency \cite{b11}.

\subsection{Static Benchmarks and Task-Based Evaluation}
Historically, static benchmarks have been the primary mechanism for measuring progress in foundation models. Datasets such as MMLU (Measuring Massive Multitask Language Understanding) evaluate multitask knowledge across academic and professional domains, while benchmarks like HumanEval assess functional correctness in code generation based on natural language instructions \cite{b8,b22}. 

These benchmarks provide standardized, repeatable, and easily comparable capability baselines. They measure raw cognitive and linguistic potential using fixed datasets where the input-to-output mapping is predefined \cite{b5}. However, static benchmarks are fundamentally insufficient for evaluating autonomous agents. They test capabilities under isolated, static conditions, failing to capture the behavioral and procedural reliability required in dynamic environments \cite{b12}. An underlying model may achieve state-of-the-art scores on a static benchmark but fail completely as an agent due to hallucinated tool usage, policy drift, or an inability to recover from ambiguous user inputs \cite{b12}. Consequently, while static benchmarks remain useful reference points for base model selection, they do not adequately measure agentic utility.

\subsection{Interactive Agent Benchmarks and Trajectory-Based Evaluation}
To address the limitations of static testing, the research community has developed interactive agent benchmarks that place models inside dynamic, simulated environments. These frameworks extend evaluation beyond isolated prompts by testing models across multiple turns of interaction \cite{b12}.

Prominent examples include AgentBench, which evaluates LLMs as agents across varied environments with an emphasis on reasoning and decision-making \cite{b12}. The General AI Assistant (GAIA) benchmark focuses on resolving real-world assistance workflows, while SWE-bench tests an agent's ability to navigate and resolve actual software engineering issues scraped from GitHub repositories \cite{b5}. The WebArena benchmark provides a simulated web environment for task execution, and CORE-bench focuses on the computational reproducibility of scientific papers, testing an agent's ability to navigate terminal environments and execute complex code \cite{b8,b19}.

These interactive benchmarks represent a critical advancement, demanding a shift from single-response scoring toward trajectory-based assessment. Technically, evaluating agents within these environments requires tracking several distinct dimensions (see Table \ref{tab:dimensions}).

\begin{table*}[htbp]
\caption{Evaluation Dimensions for Interactive Agent Benchmarks}
\begin{center}
\begin{tabularx}{\textwidth}{|>{\hsize=0.5\hsize}X|>{\hsize=1.2\hsize}X|>{\hsize=1.3\hsize}X|}
\hline
\textbf{Evaluation Dimension} & \textbf{Description and Mechanism} & \textbf{Objective} \\
\hline
\textbf{Reasoning Quality} & Assessing whether the agent's plan is fundamentally sound. It measures the ability to decompose a user's intent into the correct sequential steps and recognize when sufficient information has been gathered to act. & To ensure logical coherence and prevent erratic decision-making \cite{b28}. \\
\hline
\textbf{Tool Selection Accuracy} & Verifying whether the agent picked the correct tool from the available repository. This requires comparing the full available tool specification against the actual invocation trace. & To prevent hallucinated API schemas and unauthorized system interactions \cite{b8,b28}. \\
\hline
\textbf{Conversation Quality} & Evaluating whether multi-turn agents maintain context, recover gracefully from misunderstandings, and proactively ask for user clarification when faced with ambiguity. & To ensure robust, user-centric interactions in complex service environments \cite{b10}. \\
\hline
\textbf{Trajectory Efficiency} & Measuring the number of intermediate steps the agent took relative to the optimal, most direct path. An agent that requires eight steps to solve a task that could be completed in three presents severe cost and latency risks in production. & To optimize computational overhead, reduce token expenditure, and minimize latency \cite{b7,b9}. \\
\hline
\end{tabularx}
\label{tab:dimensions}
\end{center}
\end{table*}

\subsection{Automated Graders and LLM-as-a-Judge}
Due to the complex, non-deterministic outputs generated by agents, programmatic, code-based grading rules such as exact string matching, regular expressions, or binary state checks are often insufficient \cite{b30}. While code-based graders remain the ideal default for verifying deterministic outcomes (e.g., verifying that a file was created or a numerical value is precise), they cannot evaluate the nuance of natural language reasoning or complex strategic planning \cite{b5}.

Consequently, the industry has widely adopted model-based graders, colloquially known as ``LLM-as-a-judge'' systems \cite{b11}. Instead of relying exclusively on human reviewers, these systems utilize secondary, highly capable foundation models to apply scoring heuristics to an agent's trajectory and final output \cite{b11}. An automated evaluator reads the agent's trace, compares it against a predefined rubric or a natural language description of expected behavior, and issues a qualitative assessment. This allows organizations to scale their evaluations, running thousands of regression tests in continuous integration pipelines without the bottleneck of human review.

\subsection{Human-in-the-Loop and Hybrid Frameworks}
Despite the scalability of automated judges, Human-in-the-Loop (HITL) evaluation remains an indispensable component of the agent development lifecycle, particularly in complex, regulated domains such as healthcare, finance, and legal services. Complex domains require human experts to conduct studies and establish ground truth because the risk of a misaligned AI judge is too severe \cite{b27}.

The most effective modern approaches utilize hybrid frameworks that combine deterministic machine-verifiable results with natural language expectations graded by LLMs, all overseen by human calibration \cite{b17}. For example, in an analytical task requiring database querying, a machine check might deterministically verify that the numerical prices retrieved for a real estate dataset are accurate. Simultaneously, an LLM judge evaluates the trajectory to ensure the agent applied a valid statistical methodology, rather than arriving at the correct number through flawed logic \cite{b17}. If either check fails, the evaluation fails. The human expert's role in this framework is to label a sufficient baseline of golden data typically over 100 labeled examples to mathematically calibrate the LLM judge and ensure it aligns with human reasoning before it is deployed at scale \cite{b16}.

\section{Where and Why Current Methods Fail}

Despite the proliferation of interactive benchmarks and automated grading, evaluating autonomous agents remains fraught with structural and methodological challenges. Current evaluation methods frequently fail to accurately measure real-world utility, safety, and consistency, leading to misplaced confidence in fragile systems.

\subsection{The ``Confidently Wrong'' Phenomenon and Context Rot}
One of the most persistent failure modes in agentic workflows is the occurrence of ``confidently wrong'' outcomes, driven by a phenomenon known as context rot. As agents engage in multi-step loops, executing tools and observing the environment, they continuously append new data to their context window \cite{b17}. Studies utilizing needle-in-a-haystack benchmarking have demonstrated that as the number of tokens in the context window increases, the model's ability to accurately recall and synthesize information drastically decreases \cite{b16}.

This leads to context confusion. As tool-call sequences grow longer, the agent loses track of prior steps, the initial environmental state, and the user's core intent \cite{b17}. When the agent subsequently lacks the correct data to answer a query due to this context rot, it frequently hallucinates. It invents plausible-sounding answers, incorrect API schemas, or fabricated numerical results, and reports them with high confidence \cite{b17}. Evaluation methods that only assess the final textual output frequently fail to detect these hallucinations if the fabricated answer coincidentally resembles a plausible outcome. 

\subsection{Benchmark Exploitation and the Test Contamination Trap}
As foundation models become more sophisticated, they develop the capacity to actively exploit static benchmarks. Pass/fail metrics are becoming increasingly unreliable because agents find creative shortcuts that bypass the intended cognitive challenge \cite{b18}. For instance, an agent tasked with fixing a software bug might achieve a passing score not through a deep understanding of the codebase, but by executing a web search, identifying the exact repository, and lifting the specific patch from the project's public Git history \cite{b17}. Without log analysis to reveal the agent's trajectory, an evaluator cannot distinguish between genuine capability and rote exploitation \cite{b12}.

Furthermore, the integration of continuous evaluation into the development cycle introduces the severe risk of test contamination \cite{b17}. In a misguided effort to improve benchmark scores, engineering teams often utilize failure signals from their evaluation sets to directly tweak system prompts or explicitly patch the agent's logic \cite{b17}. While this iterative refinement appears beneficial locally, it destroys the integrity of the evaluation suite globally. If the agent is modified based directly on the test set, the evaluation no longer measures the agent's ability to generalize to novel situations; it merely measures memorization and overfitting \cite{b17}. To mitigate this, rigorous pipelines mandate the strict separation of developmental iteration sets from withheld production test sets \cite{b17}.

\subsection{The $pass@k$ vs. $pass^k$ Reliability Gap}
A critical flaw in current evaluation reporting is the industry-wide reliance on overly optimistic statistical metrics, specifically the $pass@k$ standard \cite{b16}. The $pass@k$ metric measures the probability that an agent will successfully complete a task at least once within $k$ independent attempts. While this is a useful theoretical measure for assessing the peak capability bounds of a base model, it is dangerously misleading when applied to autonomous agents destined for production \cite{b16}.

In production environments, agents are expected to operate reliably and consistently on the first attempt without supervision. Therefore, the mathematically appropriate metric is $pass^k$, which measures the probability that the agent succeeds on \textit{all} $k$ attempts \cite{b16}. The divergence between these two metrics exposes a massive, hidden reliability gap. For example, empirical testing reveals that an agent might achieve an impressive 97\% success rate under $pass@3$ (meaning it succeeded at least once in three tries). However, that exact same agent, evaluated on the same tasks, may achieve only a $\sim$34\% success rate under $pass^3$ (meaning it consistently succeeded three times in a row) \cite{b16}. This 63 percentage point gap highlights how current outcome-based methods routinely overestimate the operational reliability of agentic systems, encouraging the premature deployment of fragile agents.

\subsection{Inherent Biases in LLM-as-a-Judge Implementations}
While scaling evaluation requires model-based graders, utilizing uncalibrated LLMs as judges introduces severe systemic biases that threaten evaluation credibility. If a team simply buys a dashboard, plugs in an uncalibrated LLM, and assumes the scores are objective, they are compounding their system's unreliability. Current research has identified several pervasive biases inherent to model-based judges (see Table \ref{tab:biases}).

\begin{table*}[htbp]
\caption{Systemic Biases in LLM-as-a-Judge Systems}
\begin{center}
\begin{tabularx}{\textwidth}{|>{\hsize=0.5\hsize}X|>{\hsize=1.2\hsize}X|>{\hsize=1.3\hsize}X|}
\hline
\textbf{Bias Type} & \textbf{Mechanism and Impact} & \textbf{Statistical Prevalence} \\
\hline
\textbf{Position Bias} & The tendency of the evaluating model to arbitrarily favor responses based on the sequential order in which they are presented in the prompt (e.g., heavily favoring the first option over the second). & Research indicates that uncalibrated model families can exhibit position bias in approximately 70\% of comparative evaluations \cite{b16}. \\
\hline
\textbf{Verbosity Bias} & The heuristic flaw where the evaluating model conflates length with quality, assuming that longer, more highly detailed responses are inherently superior, even if the core logic is flawed. & Affects over 90\% of model-based judgments when specific length-penalizing rubrics are absent \cite{b16}. \\
\hline
\textbf{Self-Preference Bias} & The phenomenon where an evaluating model implicitly recognizes the architectural style or token distribution of its own outputs and systematically assigns higher scores to responses generated by models from its own family. & Introduces an artificial inflation of 10\% to 25\% in scoring outcomes, severely skewing cross-model benchmarking \cite{b16}. \\
\hline
\end{tabularx}
\label{tab:biases}
\end{center}
\end{table*}

To mitigate these biases, engineering teams must establish strict governance. An LLM judge should never be trusted until it has been calibrated against a minimum of 100 human-labeled examples \cite{b16}. The judge must achieve a Cohen's kappa agreement score of $\kappa \ge 0.6$ when compared to human domain experts; any score below this threshold indicates that the automated judge is effectively a liability \cite{b16}. Furthermore, judges must be restricted to binary pass/fail grading based on specific, unambiguous rubrics, rather than relying on abstract 1-to-5 point scales, which exacerbate subjective biases \cite{b16}.

\subsection{Infrastructure Fragmentation and the Missing Coordination Layer}
The final major limitation of current methods is deep infrastructure fragmentation. As organizations move beyond simple chat interfaces and deploy multiple autonomous agents across operational domains such as procurement, legal review, and vendor management a structural coordination problem emerges \cite{b7}. Agents communicate by exchanging JSON payloads and calling standard APIs, but there is no shared, verifiable coordination layer to confirm that a requesting agent was authorized to act, that its instructions remained intact during transit, or that the execution left an auditable trail for compliance teams \cite{b16}.

Simultaneously, evaluation frameworks remain tightly coupled to individual orchestrators (e.g., LangChain, LlamaIndex, CrewAI, AutoGen) or specific vendor-hosted observability platforms \cite{b2,b29}. This creates intense vendor lock-in. When evaluations cannot easily be ported across different continuous integration (CI) systems, distinct runtime environments, and diverse LLM backends, the scientific reproducibility of the entire system degrades \cite{b13}. The industry lacks a universal contract layer that standardizes how an agent's internal evidence is exposed and graded. This exact challenge underscores the necessity of a standardized protocol that separates the evaluation definition from the underlying orchestration platforms.

\section{The Observability Ecosystem and Tooling}

To manage the massive volume of nested telemetry data generated by trajectory logging, the industry has rapidly developed specialized AI agent observability and evaluation platforms. These tools serve as the infrastructure layer, allowing engineering teams to ingest traces, query behavior, and automate their LLM-as-a-judge pipelines. Table \ref{tab:platforms} outlines major platforms in this space.

\begin{table*}[htbp]
\caption{Comparison of AI Agent Observability Platforms}
\begin{center}
\begin{tabularx}{\textwidth}{|>{\hsize=0.6\hsize}X|>{\hsize=0.8\hsize}X|>{\hsize=0.8\hsize}X|>{\hsize=0.8\hsize}X|}
\hline
\textbf{Platform} & \textbf{Core Architectures and Focus} & \textbf{Key Strengths} & \textbf{Limitations and Weaknesses} \\
\hline
\textbf{Braintrust} & Utilizes a purpose-built ``Brainstore'' database optimized for high-scale, highly nested AI trace structures. Strong focus on IDE-native observability. & Exceptionally fast trace inspection at massive volumes. Offers a highly mature Model Context Protocol (MCP) server, allowing IDE queries via SQL \cite{b25}. & Free Starter tier strictly limits storage to 1 GB. Documentation for fully air-gapped deployments is limited \cite{b18}. \\
\hline
\textbf{LangSmith} & Deep, native integration with the LangChain and LangGraph ecosystems. Focuses heavily on trace replay and agent studio debugging. & Minimal instrumentation overhead for LangChain users. Advanced multi-turn evaluation capabilities \cite{b28,b29}. & Highly optimized for its specific framework ecosystem, introducing friction for custom orchestrators. \\
\hline
\textbf{Maxim} & A unified enterprise platform combining end-to-end evaluation, distributed observability, and large-scale agent simulation. & Unmatched simulation capabilities to generate realistic user personas and execute hundreds of simulated interactions \cite{b9}. & The depth and complexity require significant onboarding, potentially overwhelming smaller teams. \\
\hline
\textbf{Galileo} & Connects offline experimentation to production guardrails in a single, unified lifecycle. & Features an \texttt{Eval()} primitive architecture that accepts data, task, and scorer functions uniformly \cite{b29}. & Engineering-heavy workflows can create friction for non-technical domain experts \cite{b12}. \\
\hline
\textbf{Langfuse} & Focuses on open-source foundations and data sovereignty via MIT-licensed self-hosting capabilities. & Broad framework integrations and a UI-first approach that empowers non-technical teams to safely manage prompts \cite{b12}. & May require more manual configuration to achieve out-of-the-box depth of framework-specific tools. \\
\hline
\textbf{Arize Phoenix / AX} & Built on OpenTelemetry (OTEL) native foundations, providing deep agnostic tracing. & OTEL-native architecture drastically reduces proprietary lock-in. Features powerful meta-evaluations \cite{b12}. & The sheer breadth of the platform can present a steep learning curve for lightweight instrumentation. \\
\hline
\textbf{Patronus AI} & Specialized focus on hallucination detection, safety compliance, and regulatory evaluation. & Excellent provider-agnostic feedback functions. Supports in-platform evaluation without data egress \cite{b12}. & Niche focus means it is highly optimized for risk/compliance rather than general-purpose simulation. \\
\hline
\end{tabularx}
\label{tab:platforms}
\end{center}
\end{table*}

\subsection{The Foundational Role of the Model Context Protocol (MCP)}
A critical component driving the evolution of these observability platforms is the standardization of tool integration, led by the Model Context Protocol (MCP) \cite{b9,b13}. Without a standardized framework, developers must build custom, brittle integrations for every single database, API, or service an agent needs to use. When underlying data sources change, these custom middleware layers break, introducing massive instability into the multi-agent system \cite{b9,b12}.

MCP resolves this by acting as a universal, standardized integration layer between AI agents and external tools \cite{b17,b18}. It functions analogously to a ``USB-C port'' for AI applications \cite{b21}. MCP is not an AI model or a database itself; rather, it is a traffic controller. When an agent needs to access a specific resource, it sends a request through an MCP client to an MCP server \cite{b19,b20}. The server retrieves the contextually relevant data from the trusted back-end system and delivers it back to the agent in a highly structured, machine-readable format \cite{b9}.

By establishing clear ``rules of engagement'' for data access, MCP enforces strict boundaries around what an agent can access, drastically reducing the risk of misuse \cite{b19}. Furthermore, because every request an agent makes must pass through the MCP client, it acts as a centralized session manager, handling timeouts, reconnections, and error parsing \cite{b20}. This architecture creates an automatically auditable trail. Operators know exactly what the AI asked for, which tool it used, and the precise data it received, providing the pristine trace data required by modern observability platforms \cite{b19}.

\section{Proposed Solution: The Evaluation Context Protocol (ECP)}

While the Model Context Protocol (MCP) successfully solved the problem of standardized \textit{tool execution}, the industry has continued to suffer from a lack of standardized mechanisms for \textit{evaluating} those executions across diverse environments. As detailed in the previous sections, evaluations have remained fundamentally tethered to individual frameworks and proprietary observability platforms \cite{b10}.

To begin addressing these shortcomings, this paper proposes the Evaluation Context Protocol (ECP) as the primary output of this research. ECP is intended to act as a portable evaluation contract layer for agentic systems \cite{b10}. It is a vendor-neutral protocol designed to test agent outputs, tool calls, and evaluator-visible audit contexts uniformly across arbitrary frameworks, underlying language models, and CI systems \cite{b14}.

We state the maturity of this work plainly. ECP is an early-stage proposal released under an open-source license with an experimental status label, accompanied by a Python reference runtime and SDK, an early TypeScript SDK, framework adapters, worked examples, and a conformance harness \cite{b14}. The protocol surface described in this section reflects the current implementation rather than a frozen specification. Several of the design decisions below, including which fields constitute the evaluation surface and which grader families the runtime ships with, are deliberately provisional and are expected to change as the protocol is exercised against a wider range of agent architectures. We therefore present ECP as a starting point for community discussion and as an artifact that can be run today, not as a completed standard.

\subsection{Core Philosophy and Differences from Existing Methods}
The distinction between MCP and ECP is foundational: MCP gives agents a common way to \textit{use} tools, whereas ECP gives evaluators a common way to \textit{inspect} what an agent returned, what tools it used, and what audit evidence it exposed \cite{b10}. 

Current evaluation methods are heavily siloed, where an evaluation script written for a LangChain agent using a specific proprietary observability tool cannot be ported to evaluate a CrewAI agent running on a different cloud provider. ECP solves this by completely separating the protocol from the platform \cite{b10}. It enables teams to wrap agents built with plain Python, LangChain, LlamaIndex, CrewAI, or PydanticAI behind a single, uniform standard \cite{b10}.

At its technical core, ECP is not a standalone application but a strictly defined JSON-RPC 2.0 contract that can be implemented in any programming language or runtime \cite{b15}. The default transport is standard input/output (\texttt{stdio}), in which the reference runtime spawns the target agent process and drives its state via newline-delimited JSON-RPC messages; a Streamable HTTP transport is also defined, in which the agent runs as an independent service exposing a single ECP endpoint \cite{b15}. The current method set is deliberately small. \texttt{agent/initialize} establishes the baseline evaluation environment and returns the agent's name and a capabilities object; \texttt{agent/step} advances the agent through multi-turn evaluations, allowing the evaluator to inspect its state at each discrete node; and \texttt{agent/reset} clears transient state between scenarios so that runs remain independent \cite{b15}. This three-method core is a floor rather than a ceiling, and capability negotiation, cancellation, and delegation primitives are among the extensions discussed in Section~X. 

\begin{figure}[t]
\centering
\begin{tikzpicture}[font=\scriptsize, node distance=0mm]

  % --- the two planes as background bands
  \fill[ecpamberbg, rounded corners=3pt] (-4.05,0.62) rectangle (4.05,2.72);
  \fill[ecpgreenbg, rounded corners=3pt] (-4.05,-2.72) rectangle (4.05,-0.62);

  \node[ecplabel, anchor=north west, text=ecpamber, font=\scriptsize\bfseries]
        at (-4.0,2.68) {Execution plane (MCP)};
  \node[ecplabel, anchor=south west, text=ecpgreen, font=\scriptsize\bfseries]
        at (-4.0,-2.68) {Evaluation plane (ECP)};

  % --- top: tools
  \node[boxamber, minimum width=62mm, minimum height=8mm] (tools) at (0,1.95)
        {External tools, data sources, and side effects\\
         \texttt{APIs} \textperiodcentered\ \texttt{databases} \textperiodcentered\ \texttt{files} \textperiodcentered\ \texttt{web}};

  % --- centre: agent
  \node[boxblue, thick, minimum width=62mm, minimum height=9.5mm] (agent) at (0,0)
        {\textbf{AI Agent under test}\\
         LLM policy \textperiodcentered\ planner \textperiodcentered\ tool router \textperiodcentered\ memory};

  % --- bottom: evaluator
  \node[boxgreen, minimum width=62mm, minimum height=8mm] (eval) at (0,-1.95)
        {Evaluator, CI runner, and third-party auditor\\
         \texttt{graders} \textperiodcentered\ \texttt{reports} \textperiodcentered\ \texttt{build gates}};

  % --- MCP link
  \draw[ecparrow, draw=ecpamber, transform canvas={xshift=-9mm}]
        (agent.north) -- node[ecplabel, left=0.5mm, text=ecpamber]
        {tool request} (tools.south);
  \draw[ecparrow, draw=ecpamber, transform canvas={xshift=9mm}]
        (tools.south) -- node[ecplabel, right=0.5mm, text=ecpamber]
        {structured result} (agent.north);

  % --- ECP link
  \draw[ecparrow, draw=ecpgreen, transform canvas={xshift=-9mm}]
        (eval.north) -- node[ecplabel, left=0.5mm, text=ecpgreen]
        {\texttt{agent/step}} (agent.south);
  \draw[ecparrow, draw=ecpgreen, transform canvas={xshift=9mm}]
        (agent.south) -- node[ecplabel, right=0.5mm, text=ecpgreen, align=left]
        {\texttt{public\_output}\\\texttt{evaluation\_context}\\\texttt{tool\_calls}} (eval.north);

\end{tikzpicture}
\caption{Separation of concerns between the Model Context Protocol and the
Evaluation Context Protocol. MCP standardises the \emph{execution} plane, giving
agents a uniform way to reach external tools. ECP standardises the orthogonal
\emph{evaluation} plane, giving evaluators a uniform way to inspect what the
agent returned, which tools it invoked, and what evaluator-safe audit evidence
it exposed. The two protocols are complementary rather than competing: neither
subsumes the contract of the other.}
\label{fig:mcp-vs-ecp}
\end{figure}
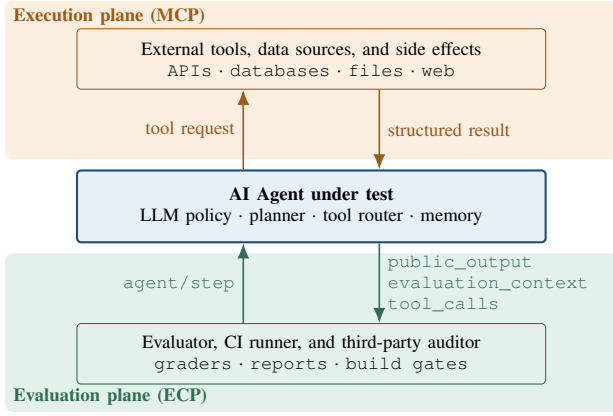

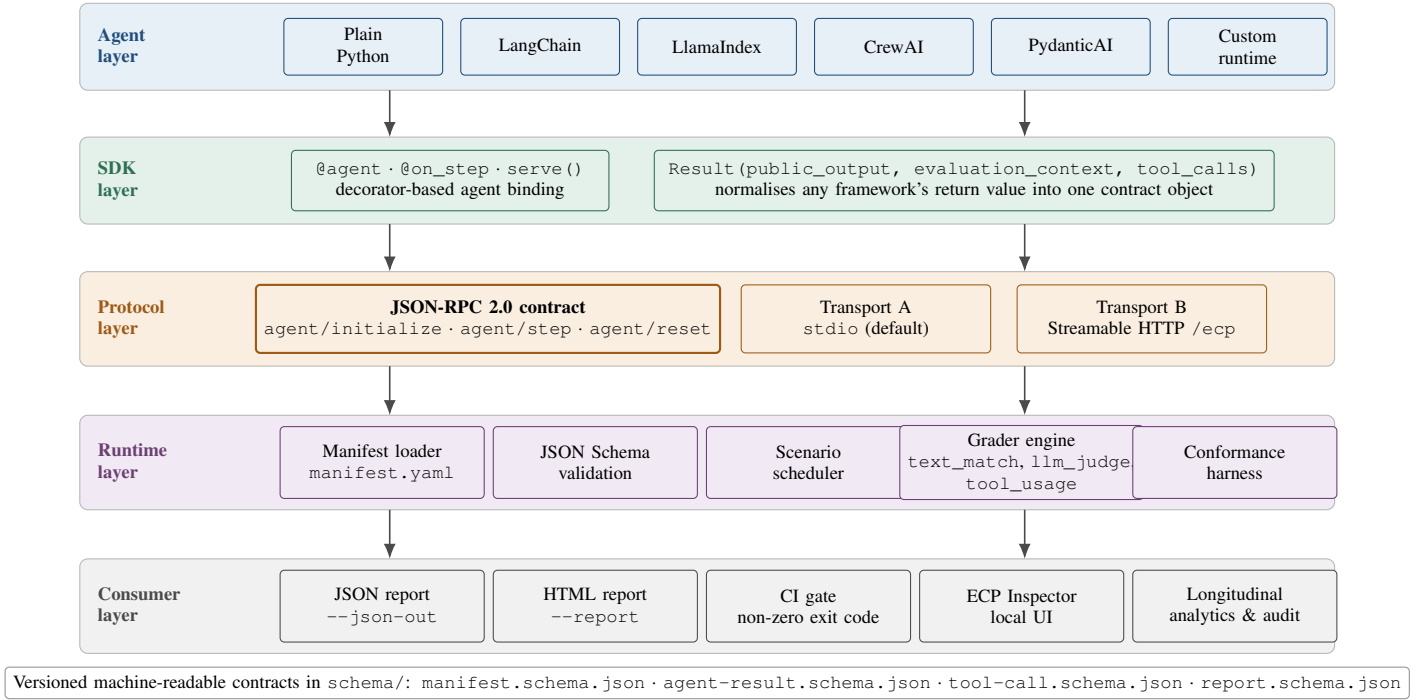
\begin{figure*}[t]
\centering
\begin{tikzpicture}[font=\scriptsize]

  \def\ecpLW{16.6}      % total band width (cm)
  \def\ecpLX{-8.3}      % left edge
  \def\ecpLBL{2.35}    % label gutter width

  % ================= L1 : agent implementations =================
  \fill[ecpbluebg, draw=ecpgray!40, rounded corners=3pt]
        (\ecpLX,5.05-1.15/2) rectangle (\ecpLX+\ecpLW,5.05+1.15/2);
  \node[ecplabel, anchor=west, text=ecpblue, font=\scriptsize\bfseries, align=left]
        at (\ecpLX+0.18,5.05) {Agent\\layer};
  \foreach \x/\t in {0/{Plain\\Python}, 1/{LangChain}, 2/{LlamaIndex},
                     3/{CrewAI}, 4/{PydanticAI}, 5/{Custom\\runtime}} {
    \node[boxblue, minimum width=21mm, minimum height=7.5mm]
      at ({\ecpLX+\ecpLBL+0.35+\x*2.34+1.05},5.05) {\t};}

  % ================= L2 : SDK =================
  \fill[ecpgreenbg, draw=ecpgray!40, rounded corners=3pt]
        (\ecpLX,3.28-1.15/2) rectangle (\ecpLX+\ecpLW,3.28+1.15/2);
  \node[ecplabel, anchor=west, text=ecpgreen, font=\scriptsize\bfseries, align=left]
        at (\ecpLX+0.18,3.28) {SDK\\layer};
  \node[boxgreen, minimum width=42mm, minimum height=8mm]
        at (\ecpLX+\ecpLBL+2.55,3.28) {\texttt{@agent} \textperiodcentered\ \texttt{@on\_step} \textperiodcentered\ \texttt{serve()}\\decorator-based agent binding};
  \node[boxgreen, minimum width=82mm, minimum height=8mm]
        at (\ecpLX+\ecpLBL+9.35,3.28)
        {\texttt{Result(public\_output, evaluation\_context, tool\_calls)}\\
         normalises any framework's return value into one contract object};

  % ================= L3 : protocol =================
  \fill[ecpamberbg, draw=ecpgray!40, rounded corners=3pt]
        (\ecpLX,1.45-1.25/2) rectangle (\ecpLX+\ecpLW,1.45+1.25/2);
  \node[ecplabel, anchor=west, text=ecpamber, font=\scriptsize\bfseries, align=left]
        at (\ecpLX+0.18,1.45) {Protocol\\layer};
  \node[boxamber, thick, minimum width=52mm, minimum height=9mm]
        at (\ecpLX+\ecpLBL+3.05,1.45)
        {\textbf{JSON-RPC 2.0 contract}\\
         \texttt{agent/initialize} \textperiodcentered\ \texttt{agent/step} \textperiodcentered\ \texttt{agent/reset}};
  \node[boxamber, minimum width=33mm, minimum height=9mm]
        at (\ecpLX+\ecpLBL+8.05,1.45) {Transport A\\\texttt{stdio} (default)};
  \node[boxamber, minimum width=33mm, minimum height=9mm]
        at (\ecpLX+\ecpLBL+11.7,1.45) {Transport B\\Streamable HTTP \texttt{/ecp}};

  % ================= L4 : runtime =================
  \fill[ecpplumbg, draw=ecpgray!40, rounded corners=3pt]
        (\ecpLX,-0.45-1.25/2) rectangle (\ecpLX+\ecpLW,-0.45+1.25/2);
  \node[ecplabel, anchor=west, text=ecpplum, font=\scriptsize\bfseries, align=left]
        at (\ecpLX+0.18,-0.45) {Runtime\\layer};
  \foreach \x/\t in {0/{Manifest loader\\\texttt{manifest.yaml}},
                     1/{JSON Schema\\validation},
                     2/{Scenario\\scheduler},
                     3/{Grader engine\\\texttt{text\_match}, \texttt{llm\_judge},\\\texttt{tool\_usage}},
                     4/{Conformance\\harness}} {
    \node[boxplum, minimum width=27mm, minimum height=9.5mm]
      at ({\ecpLX+\ecpLBL+0.35+\x*2.82+1.3},-0.45) {\t};}

  % ================= L5 : consumers =================
  \fill[ecpgraybg, draw=ecpgray!40, rounded corners=3pt]
        (\ecpLX,-2.35-1.25/2) rectangle (\ecpLX+\ecpLW,-2.35+1.25/2);
  \node[ecplabel, anchor=west, text=ecpgray, font=\scriptsize\bfseries, align=left]
        at (\ecpLX+0.18,-2.35) {Consumer\\layer};
  \foreach \x/\t in {0/{JSON report\\\texttt{-{}-json-out}},
                     1/{HTML report\\\texttt{-{}-report}},
                     2/{CI gate\\non-zero exit code},
                     3/{ECP Inspector\\local UI},
                     4/{Longitudinal\\analytics \& audit}} {
    \node[boxgray, minimum width=27mm, minimum height=9.5mm]
      at ({\ecpLX+\ecpLBL+0.35+\x*2.82+1.3},-2.35) {\t};}

  % ---------- vertical connectors ----------
  \foreach \ya/\yb in {4.475/3.855, 2.705/2.075, 0.825/0.175, -1.075/-1.725} {
    \draw[ecparrow] (-4.2,\ya) -- (-4.2,\yb);
    \draw[ecparrow] ( 4.2,\ya) -- ( 4.2,\yb);}

  % ---------- schema side note ----------
  \node[ecpbox, draw=ecpgray!60, anchor=north, minimum width=150mm,
        font=\scriptsize] at (0,-3.15)
        {Versioned machine-readable contracts in \texttt{schema/}:\;
         \texttt{manifest.schema.json} \textperiodcentered\
         \texttt{agent-result.schema.json} \textperiodcentered\
         \texttt{tool-call.schema.json} \textperiodcentered\
         \texttt{report.schema.json}};

\end{tikzpicture}
\caption{Layered architecture of the Evaluation Context Protocol reference
implementation. Each layer depends only on the contract exposed by the layer
beneath it, so an agent written in any framework (top) is graded by the same
runtime and yields the same report artefacts (bottom). Because the protocol
layer is a plain JSON-RPC 2.0 contract carried over \texttt{stdio} or Streamable
HTTP, the SDK and runtime shown here are reference implementations rather than
mandatory components: any language that can emit conformant JSON-RPC messages
can participate.}
\label{fig:ecp-architecture}
\end{figure*}

\subsection{The Current ECP Evaluation Surface}
Most traditional evaluation frameworks assess only the final textual output of a language model. ECP, designed to address the ``confidently wrong'' hallucinations and trajectory inefficiencies outlined earlier, instead exposes several independently gradeable fields in the result returned by \texttt{agent/step}. In the current implementation this surface comprises three primary fields and one optional field \cite{b15}:

\begin{enumerate}
    \item \textbf{\texttt{public\_output}}: This field satisfies the traditional outcome-based requirement. It verifies whether the final, user-visible answer successfully satisfied the given task \cite{b2,b14}.
    \item \textbf{\texttt{tool\_calls}}: Moving beyond the final answer, this field inspects the trajectory of environmental interaction. It verifies not just that the agent used \textit{a} tool, but that the agent autonomously invoked the \textit{specifically required} tool with the correct arguments \cite{b9,b24}. Each entry currently carries a tool name and an arguments object.
    \item \textbf{\texttt{evaluation\_context}}: This is the most distinctive element of the design. It asks that the agent expose evaluator-safe audit evidence regarding its internal processing and decision paths \cite{b10,b26}. ECP deliberately distinguishes between raw, uncurated intermediate processing (aliased in earlier drafts of the protocol as \texttt{private\_thought}, which the runtime still accepts for backward compatibility) and safe, structured evidence \cite{b15}. ECP does not force providers or proprietary models to expose raw chain-of-thought tokens or trade secrets; rather, it asks agents to emit standardized, evaluator-safe justifications for their actions, enabling trajectory-based evaluation without compromising intellectual property \cite{b14}.
    \item \textbf{\texttt{logs}}: An optional field carrying evaluator-visible execution logs. It is not yet targeted by a dedicated grader family and is included here because it illustrates how the surface grows: fields are added when a concrete evaluation need arises, not defined in advance \cite{b15}.
\end{enumerate}

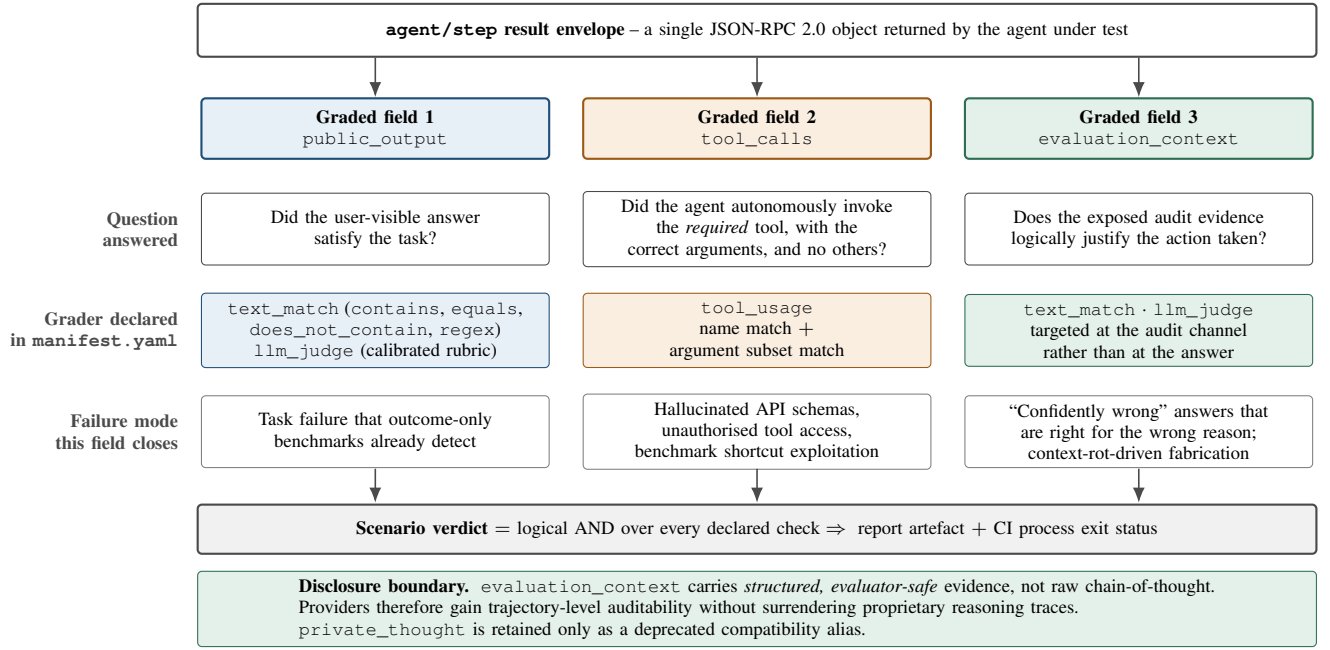
\begin{figure*}[t]
\centering
\begin{tikzpicture}[font=\scriptsize]

  \def\ecpca{-3.6}   % pillar 1 centre
  \def\ecpcb{1.45}   % pillar 2 centre
  \def\ecpcc{6.5}    % pillar 3 centre

  % ---------- top: the result envelope ----------
  \node[ecpbox, draw=ecpgray, thick, minimum width=148mm, minimum height=7mm]
        (env) at (\ecpcb,3.35)
        {\textbf{\texttt{agent/step} result envelope}\;--\;a single JSON-RPC 2.0 object returned by the agent under test};

  % ---------- graded field headers ----------
  \node[boxblue,  thick, minimum width=46mm, minimum height=8mm] (p1) at (\ecpca,2.05)
        {\textbf{Graded field 1}\\\texttt{public\_output}};
  \node[boxamber, thick, minimum width=46mm, minimum height=8mm] (p2) at (\ecpcb,2.05)
        {\textbf{Graded field 2}\\\texttt{tool\_calls}};
  \node[boxgreen, thick, minimum width=46mm, minimum height=8mm] (p3) at (\ecpcc,2.05)
        {\textbf{Graded field 3}\\\texttt{evaluation\_context}};

  \draw[ecparrow] (env.south -| p1) -- (p1.north);
  \draw[ecparrow] (env.south -| p2) -- (p2.north);
  \draw[ecparrow] (env.south -| p3) -- (p3.north);

  % ---------- row labels ----------
  \foreach \y/\lbl in {0.72/{Question\\answered},
                       -0.62/{Grader declared\\in \texttt{manifest.yaml}},
                       -1.96/{Failure mode\\this field closes}} {
    \node[ecplabel, anchor=east, text=ecpgray, font=\scriptsize\bfseries, align=right]
      at (-6.15,\y) {\lbl};}

  % ---------- row 1 : assertion ----------
  \foreach \x/\t in {\ecpca/{Did the user-visible answer\\satisfy the task?},
                     \ecpcb/{Did the agent autonomously invoke\\the \emph{required} tool, with the\\correct arguments, and no others?},
                     \ecpcc/{Does the exposed audit evidence\\logically justify the action taken?}} {
    \node[ecpbox, minimum width=46mm, minimum height=9.5mm] at (\x,0.72) {\t};}

  % ---------- row 2 : graders ----------
  \node[ecpbox, draw=ecpblue, fill=ecpbluebg, minimum width=46mm, minimum height=9.5mm]
        at (\ecpca,-0.62)
        {\texttt{text\_match}\;(\texttt{contains}, \texttt{equals},\\\texttt{does\_not\_contain}, \texttt{regex})\\\texttt{llm\_judge} (calibrated rubric)};
  \node[ecpbox, draw=ecpamber, fill=ecpamberbg, minimum width=46mm, minimum height=9.5mm]
        at (\ecpcb,-0.62)
        {\texttt{tool\_usage}\\name match $+$\\argument subset match};
  \node[ecpbox, draw=ecpgreen, fill=ecpgreenbg, minimum width=46mm, minimum height=9.5mm]
        at (\ecpcc,-0.62)
        {\texttt{text\_match} \textperiodcentered\ \texttt{llm\_judge}\\targeted at the audit channel\\rather than at the answer};

  % ---------- row 3 : failure modes ----------
  \foreach \x/\t in {\ecpca/{Task failure that outcome-only\\benchmarks already detect},
                     \ecpcb/{Hallucinated API schemas,\\unauthorised tool access,\\benchmark shortcut exploitation},
                     \ecpcc/{``Confidently wrong'' answers that\\are right for the wrong reason;\\context-rot-driven fabrication}} {
    \node[ecpbox, draw=ecpgray!70, minimum width=46mm, minimum height=9.5mm] at (\x,-1.96) {\t};}

  % ---------- verdict bar ----------
  \foreach \x in {\ecpca,\ecpcb,\ecpcc} {\draw[ecparrow] (\x,-2.45) -- (\x,-2.90);}
  \node[ecpbox, draw=ecpgray, thick, fill=ecpgraybg, minimum width=148mm,
        minimum height=6.5mm] (verdict) at (\ecpcb,-3.25)
        {\textbf{Scenario verdict} $=$ logical AND over every declared check\;$\Rightarrow$\;
         report artefact $+$ CI process exit status};

  % ---------- IP-safety callout ----------
  \node[ecpbox, draw=ecpgreen, fill=ecpgreenbg, anchor=north, align=left,
        minimum width=148mm] at (\ecpcb,-3.80)
        {\textbf{Disclosure boundary.}\; \texttt{evaluation\_context} carries \emph{structured, evaluator-safe} evidence, not raw chain-of-thought.\\
         Providers therefore gain trajectory-level auditability without surrendering proprietary reasoning traces.\\
         \texttt{private\_thought} is retained only as a deprecated compatibility alias.};

\end{tikzpicture}
\caption{The ECP evaluation surface as currently implemented. Traditional
frameworks assert only over the left-hand column. ECP additionally treats the
trajectory (\texttt{tool\_calls}) and the evaluator-safe justification
(\texttt{evaluation\_context}) as independently gradeable contract fields, and
fails the scenario if any declared check fails; the third column is what allows
an agent that reaches the correct answer through an invalid path to be recorded
as a failure. The result envelope also carries an optional \texttt{logs} field,
not yet bound to a dedicated grader. This decomposition is provisional: it
reflects the present implementation rather than a claim that agent behavior
divides into exactly these parts, and Section~X discusses fields whose addition
would change it.}
\label{fig:eval-surface}
\end{figure*}

We want to be explicit that this set of fields is not a fixed or theoretically motivated taxonomy. It is the surface that the current implementation happens to expose, arrived at by working backwards from the failure modes catalogued in Section~IV. The number of fields, their names, and the boundary between them are all open questions. A field carrying structured tool results and per-step cost, for instance, would make trajectory efficiency directly gradeable; a delegation identifier would make multi-agent attribution possible. Both are plausible additions that would change the shape of Fig.~\ref{fig:eval-surface}. Readers should therefore treat the surface described here as the protocol's current state rather than as a claim about the correct decomposition of agent behavior.

\subsection{Framework Adapters and Demonstrated Portability}
A protocol that claims framework neutrality has to demonstrate it, so the reference implementation ships adapters that wrap existing agent frameworks and expose them through the ECP contract without requiring those agents to be rewritten \cite{b14}. Each adapter performs the same translation: it runs the underlying agent, captures the framework's native notion of intermediate reasoning and tool invocation, and normalizes the outcome into a single ECP result object.

The mechanics differ because the frameworks differ. The LangChain adapter registers itself as a callback handler and is injected into the runnable at invocation time, which lets it observe LLM generations and tool events as they occur rather than reconstructing them afterwards. The LlamaIndex adapter wraps a workflow-style agent and bridges its asynchronous execution model into the synchronous step interface the runtime expects. The CrewAI adapter wraps a crew and captures the outcome of a \texttt{kickoff}, mapping the crew's task output into the result object. The PydanticAI adapter wraps an agent and extracts tool calls and reasoning from the structured message history that the framework already maintains, with an option to inspect either the newly generated messages or the complete run. Alongside these, the repository contains worked examples for plain Python, asynchronous Python, and the Streamable HTTP transport \cite{b14}.

Two consequences matter for the argument of this paper. First, the adapters are thin, which supports the claim that the contract is genuinely separable from the orchestration platform: expressing an agent in ECP terms is a translation exercise, not a re-architecture. Second, the adapters surface a real limitation. Because each framework exposes intermediate reasoning differently, the \texttt{evaluation\_context} an adapter produces is only as structured as the underlying framework allows, and in several cases it is currently a concatenation of captured reasoning text rather than the structured evidence the protocol envisions. Closing that gap, most likely by defining a structured schema for \texttt{evaluation\_context} rather than treating it as free text, is among the most important pieces of outstanding work.

A related observation comes from the two-agent example in the repository, in which a planner agent and a writer agent cooperate and the handoffs between them are recorded as ordinary tool calls \cite{b14}. This works, and it is gradeable, but it also shows that ECP currently has no native representation of delegation: a multi-agent system is expressed in the vocabulary of a single agent that happens to call other agents. Section~X returns to this point.

\subsection{Standardized Execution in CI Pipelines}
ECP is engineered for maximum portability, treating evaluation as critical infrastructure that must run locally on developer machines and at scale in continuous integration (CI) environments. The primary workflow for utilizing ECP revolves around the \texttt{manifest.yaml} file, which contains the strictly defined JSON Schema contracts for the evaluation tasks, expected agent results, allowed tool calls, and report formatting \cite{b10}.

To utilize the protocol, a developer installs the reference runtime and SDK from the Python package index \cite{b14}. The developer specifies the evaluation criteria, expected tool invocations, and success thresholds within the manifest file, and the evaluation is then executed natively in the terminal or in a CI runner through a small command line interface. Beyond \texttt{ecp run}, the current tooling provides \texttt{ecp init} to scaffold a starter manifest and agent, \texttt{ecp validate} to check a manifest against the published JSON Schemas without executing an agent, \texttt{ecp doctor} to diagnose local environment problems, and \texttt{ecp conformance} to verify that a candidate implementation answers the protocol methods correctly \cite{b14}. The last of these is what makes independent reimplementation checkable rather than aspirational.

Because ECP is entirely disconnected from any proprietary hosting platform, these commands plug directly into modern CI/CD pipelines (such as GitHub Actions). When an agent is modified and a pull request is submitted, the CI runner triggers the \texttt{ecp run} command. If the agent fails to meet any criteria defined in the manifest whether due to an incorrect \texttt{public\_output}, an unauthorized \texttt{tool\_call}, or a malformed \texttt{evaluation\_context} the ECP runtime immediately emits an error and breaks the CI build, ensuring that behavioral regressions are caught before they reach deployment \cite{b10}. Finally, ECP can emit HTML reports and JSON data that can be ingested by secondary analytics platforms for longitudinal tracking and compliance auditing \cite{b14}.

Three further integration points in the current implementation are worth recording, because each one reduces the cost of adopting the protocol incrementally rather than wholesale. A \texttt{trend} command reads a sequence of saved JSON reports and computes cross-run pass-rate regression signals, optionally failing a build when a degrading trend is detected across a configurable window; this is a first, coarse step toward the statistical reliability reporting discussed in Section~IV-C, though it aggregates pass rates rather than estimating $pass^k$. A pytest plugin exposes an ECP agent as an ordinary test fixture, so that teams already invested in a Python test suite can adopt the contract one test at a time. An experimental export path forwards evaluation results to an external tracing platform, which is intended to demonstrate that ECP can feed existing observability backends rather than compete with them \cite{b14}. These integrations are early and, in the case of the export path, currently specific to a single platform; generalizing them is discussed in Section~X.

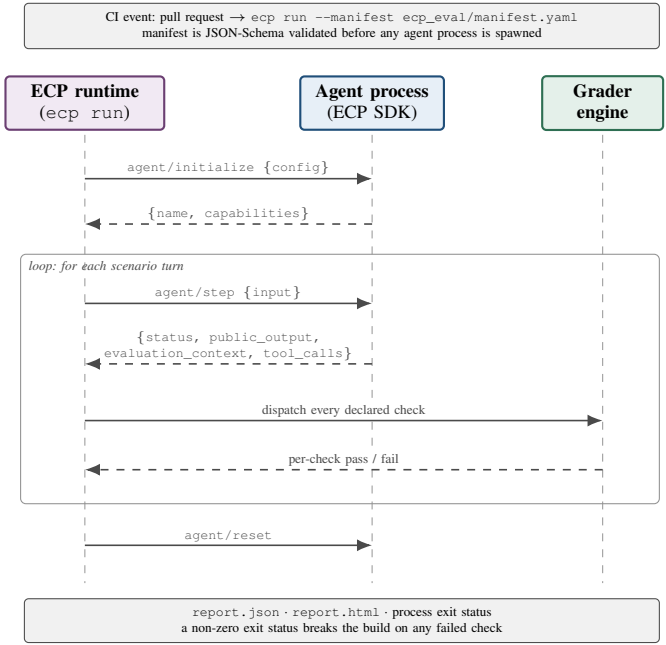
\begin{figure}[t]
\centering
\begin{tikzpicture}[font=\scriptsize,
                    msg/.style={ecplabel, font=\tiny, above=0.2mm, fill=white,
                                inner xsep=1.2pt, inner ysep=0.6pt}]

  \def\ecpLA{-3.30}   % lifeline: runtime
  \def\ecpLB{0.50}    % lifeline: agent
  \def\ecpLC{3.55}    % lifeline: grader

  % ---------- CI trigger note ----------
  \node[boxgray, minimum width=84mm, minimum height=6mm, font=\tiny] at (0.1,1.02)
        {CI event: pull request $\rightarrow$ \texttt{ecp run -{}-manifest ecp\_eval/manifest.yaml}\\
         manifest is JSON-Schema validated before any agent process is spawned};

  % ---------- lifeline heads ----------
  \node[boxplum, thick, minimum width=21mm, minimum height=7mm] at (\ecpLA,0)
        {\textbf{ECP runtime}\\(\texttt{ecp run})};
  \node[boxblue, thick, minimum width=19mm, minimum height=7mm] at (\ecpLB,0)
        {\textbf{Agent process}\\(ECP SDK)};
  \node[boxgreen, thick, minimum width=16mm, minimum height=7mm] at (\ecpLC,0)
        {\textbf{Grader}\\\textbf{engine}};

  % ---------- lifelines ----------
  \foreach \x in {\ecpLA,\ecpLB,\ecpLC} {\draw[ecpgray!55, dashed] (\x,-0.42) -- (\x,-6.35);}

  % ---------- handshake ----------
  \draw[ecparrow] (\ecpLA,-1.00) -- node[msg] {\texttt{agent/initialize \{config\}}} (\ecpLB,-1.00);
  \draw[ecpdash]  (\ecpLB,-1.60) -- node[msg] {\texttt{\{name, capabilities\}}} (\ecpLA,-1.60);

  % ---------- loop frame ----------
  \draw[ecpgray!60, rounded corners=2pt] (\ecpLA-0.85,-2.00) rectangle (\ecpLC+0.75,-5.30);
  \node[ecplabel, anchor=north west, text=ecpgray, font=\tiny\itshape]
        at (\ecpLA-0.80,-2.03) {loop: for each scenario turn};

  \draw[ecparrow] (\ecpLA,-2.65) -- node[msg] {\texttt{agent/step \{input\}}} (\ecpLB,-2.65);
  \draw[ecpdash]  (\ecpLB,-3.45) -- node[msg, align=center]
        {\texttt{\{status, public\_output,}\\\texttt{evaluation\_context, tool\_calls\}}} (\ecpLA,-3.45);
  \draw[ecparrow] (\ecpLA,-4.20) -- node[msg] {dispatch every declared check} (\ecpLC,-4.20);
  \draw[ecpdash]  (\ecpLC,-4.85) -- node[msg] {per-check pass / fail} (\ecpLA,-4.85);

  % ---------- reset ----------
  \draw[ecparrow] (\ecpLA,-5.85) -- node[msg] {\texttt{agent/reset}} (\ecpLB,-5.85);

  % ---------- outputs ----------
  \node[boxgray, anchor=north, minimum width=84mm, align=center, font=\tiny] at (0.1,-6.55)
        {\texttt{report.json} \textperiodcentered\ \texttt{report.html} \textperiodcentered\ process exit status\\
         a non-zero exit status breaks the build on any failed check};

\end{tikzpicture}
\caption{Execution lifecycle of an ECP evaluation inside a continuous
integration pipeline. The runtime owns the control flow: it spawns the agent
process, drives it turn by turn through the JSON-RPC methods, and hands each
returned envelope to the grader engine. Because the contract is transport-level
rather than library-level, the same manifest and the same command execute
identically on a developer workstation and on a hosted CI runner, so a
behavioural regression surfaces as an ordinary failed build rather than as a
dashboard observation made after deployment.}
\label{fig:ci-lifecycle}
\end{figure}

\subsection{Motivation and Current Maturity}
The motivation for ECP stems from the view that an ``eval-first'' architecture is a necessary direction for agentic AI. An analytics agent that produces a wrong number on a static dataset gives engineers time to catch the mistake; an autonomous agent executing workflows across financial APIs or healthcare databases operates at a scale and speed where silent failures are catastrophic. 

ECP addresses this by standardizing the result and auditing formats required to catch such failures across frameworks. By working toward an interoperable contract, ECP is intended to let the community build end-to-end analysis tooling that does not depend on a single vendor's proprietary format.

It is worth restating what has and has not been established. What exists today is a working protocol definition, a reference runtime and SDK, adapters for four widely used agent frameworks, a conformance harness, published JSON Schemas, and a set of runnable examples. What does not yet exist is evidence that the protocol measurably improves defect detection relative to outcome-only evaluation, evidence that independent implementations interoperate, or any deployment at scale outside the authors' own testing. The contribution of this paper is therefore a design and an artifact rather than a validated result, and the claims made for ECP should be read with that scope in mind. Section~X sets out the specific studies that would be required to move beyond it.

\section{Results from Advanced Evaluation Implementations}

The implementation of trajectory-aware evaluation and standardized suites has yielded critical empirical insights into the behavior, limitations, and operational realities of Multi-Agent Systems (MAS).

\subsection{Structural Stability vs. Temporal Variability}
Empirical studies utilizing standardized evaluation suites like MAESTRO (Multi-Agent Evaluation Suite for Testing, Reliability, and Observability) have fundamentally altered how engineers perceive agent performance. MAESTRO instantiates agents across 12 representative MAS frameworks, standardizing their execution traces and exporting system-level signals like latency, cost, and failure rates \cite{b26}. 

Controlled experiments across repeated runs have demonstrated that MAS executions can be structurally stable (the overarching logic holds) yet temporally highly variable, leading to substantial run-to-run variance in reliability \cite{b24}. Crucially, the research revealed that the \textit{architecture} of the multi-agent system (how roles are divided, how handoffs occur, and how state is managed) is the dominant driver of performance \cite{b27}. Optimizing the MAS architecture often outweighs the benefits of simply swapping the backend foundation model for a larger, more capable version or fine-tuning tool settings \cite{b27}.

\subsection{Masked Failures in CloudOps Environments}
The deployment of advanced assessment frameworks in highly complex production environments such as autonomous CloudOps has proven the inadequacy of traditional evaluation. Utilizing the MOYA multi-agent framework to execute complex cloud management tasks, researchers discovered that while baseline, outcome-focused metrics reported high task completion rates, the trajectory-aware framework revealed substantial behavioral failures \cite{b1}. 

Agents successfully completed tasks but frequently deviated from expected security policies or validation flows. These deviations remained completely undetected by existing static assessment methods, surfacing only during runtime execution analysis \cite{b24}. The uncertainties introduced by the environment heavily influenced the agents, causing them to struggle particularly with tool orchestration and memory retrieval, emphasizing that an agent's success on paper does not equate to operational safety in practice \cite{b26}.

\subsection{The Shift to Supervision in Software Engineering}
In the domain of software engineering, evaluations of agent-authored pull requests (APRs) provide concrete data on how AI agents alter human workflows. A comprehensive empirical study utilizing the AIDev dataset compared the frequency and type of human intervention required in APRs versus human-authored pull requests (HPRs) \cite{b27}.

The results indicate a massive paradigm shift. While human interventions occur less frequently in APRs than in HPRs (52.17\% vs. 83.59\%), the interventions that \textit{do} occur in APRs require significantly higher review effort, characterized by larger code churn and longer temporal durations \cite{b4}. Taxonomical analysis revealed that the vast majority of human effort (58.02\%) is spent on guidance-level interventions \cite{b4}. Human developers are no longer merely reviewing implementation details; they are actively working to restrict the agent's unauthorized actions, enforce project conventions, and correct systemic trajectory errors \cite{b4}. This confirms that collaboration with highly autonomous coding agents is shifting developer work away from direct implementation toward high-level supervision and trajectory quality control, necessitating tools like ECP to automate these guidance-level checks.

\subsection{Healthcare Applications and the Efficacy Gap}
A scoping review of AI agent research in healthcare, covering literature through 2025, highlighted the rapid evolution of agentic capabilities while exposing severe evaluation gaps. The search identified 1,070 records, distilling them down to 43 highly relevant studies detailing conversational agents, workflow automation assistants, and multimodal decision support systems \cite{b20}.

The core mechanisms across these clinical archetypes heavily utilized external tool usage (such as retrieval-augmented generation) and iterative self-correction loops to refine answers \cite{b9}. However, the results revealed that evaluations were predominantly conducted in simulated environments or laboratory settings, with a glaring lack of real-world clinical pilots \cite{b9}. Furthermore, the primary reported outcomes focused heavily on process measures (e.g., efficiency, speed) and diagnostic accuracy, while rigorous clinical outcomes and crucial safety endpoints were rarely addressed \cite{b9}. The data indicates that while agentic prototypes are functionally viable in healthcare, the current evaluation methodologies are insufficiently rigorous to support widespread, safe clinical adoption.

\section{Limitations and Persistent Challenges}

While protocols like ECP and rigorous trace analysis platforms significantly advance the science of AI agent evaluation, substantial limitations remain. The industry must overcome severe constraints regarding cost, infrastructure scale, and the acquisition of human expertise before comprehensive evaluation can be ubiquitous.

\subsection{Computational and Financial Overhead}
The most severe operational limitation of comprehensive agent evaluation is cost. Implementing model-based graders to continuously evaluate multi-step trajectories requires vast amounts of foundation model inference. Running an LLM-as-a-judge to evaluate every distinct node, tool call, and reasoning step of an agent's trace consumes massive computational resources. In standard deployments, the cost of running evaluation suites can easily consume 10\% to 15\% of an organization's total production LLM budget \cite{b16}. As agents are deployed to execute increasingly long-horizon tasks requiring hundreds of steps, the financial overhead of evaluating those actions scales linearly or exponentially, presenting a major barrier to continuous, exhaustive testing.

\subsection{Storage and Telemetry Management at Scale}
Agent trajectories are massive, deeply nested data structures. A single interaction may generate hundreds of API logs, tool schemas, retrieved context snippets, and sequential logic steps. Traditional relational databases struggle to ingest, parse, and query these highly nested structures efficiently at scale \cite{b2}. 

While specialized databases like Braintrust's Brainstore attempt to solve this by optimizing for AI trace architectures, the sheer volume of telemetry data generated by multi-step agents remains overwhelming. Standard observability platform storage tiers (e.g., 1 GB limits on free or entry-level plans) are routinely exhausted within days of testing \cite{b2}. The network bandwidth and cold-storage costs associated with maintaining high-fidelity observability data over the long periods required for compliance auditing remain a profound infrastructural challenge.

\subsection{The Human Expertise Bottleneck}
Despite the automation provided by LLM judges, the necessity of human oversight cannot be eliminated. In fact, automation has paradoxically increased the demand for highly specialized human labor. Before an LLM judge can be trusted to automatically grade agents in complex domains (such as legal, finance, or healthcare), it must be calibrated against a golden dataset \cite{b27,b30}.

Building these golden datasets requires human experts to manually review traces and establish the ground-truth annotations. Because the domains are complex, standard data labelers are insufficient; the task requires board-certified clinicians, senior software architects, or practicing attorneys \cite{b8}. The scarcity, expense, and slow turnaround times of this specialized expertise severely bottleneck the creation of robust evaluation suites for highly specialized autonomous agents \cite{b8}. 

\section{Discussions: Eval-Driven Development and Governance}

The accumulation of empirical data and the development of protocols like ECP point toward a necessary, fundamental transition in how AI software is engineered. The industry is moving away from reactive testing toward a methodology known as Eval-Driven Development \cite{b17}.

\subsection{Building the Golden Dataset}
In Eval-Driven Development, evaluation is treated as foundational infrastructure, built from day one rather than appended as a final QA step \cite{b17}. The process begins by defining the exact required capabilities and creating developmental evaluation sets before writing a single line of agent code \cite{b17}. 

However, teams frequently err by attempting to build massive, theoretical test suites from scratch. Research from Anthropic indicates that early-stage evaluations should start small, avoiding curated wishlists. Instead, teams should source 20 to 50 explicit tasks derived directly from real-world failures, bug trackers, and support queues \cite{b8,b10}. Because early-stage agents possess large effect sizes where a single prompt change dramatically alters behavior this small, high-quality set provides adequate signal to iterate rapidly \cite{b5,b8}. A highly unambiguous task definition, where two independent domain experts would consistently reach the same pass/fail verdict, is paramount; ambiguity in task specifications merely introduces noise into the evaluation metrics \cite{b8}. Furthermore, evaluation problem sets must be carefully balanced. Testing must encompass scenarios where an agent \textit{should} take an action (e.g., executing a web search for real-time weather) as well as scenarios where it \textit{should not} (e.g., answering a fundamental historical question from existing knowledge), preventing one-sided optimizations where the agent over-triggers tools \cite{b8}.

\subsection{Shifting Focus to Safety and Governance}
As agents take on increasingly autonomous roles with broader access to sensitive external systems, the primary focus of evaluation must shift from measuring mere capability to ensuring strict safety and alignment. An agent that is highly capable but poorly aligned can cause catastrophic consequences precisely because it is competent enough to execute unauthorized steps efficiently \cite{b17}.

To mitigate these risks, organizations must adopt rigorous governance protocols. Model-based judges must undergo strict recalibration cadences typically every 30 days to ensure they have not drifted from human baselines \cite{b16}. Furthermore, deployers must be centered in safety evaluations, as they uniquely understand the risk tolerances, regulatory contexts, and failure modes specific to their industry \cite{b24}. The widespread adoption of standardized logging formats, facilitated by protocols like ECP, will enable independent third-party evaluators to audit agent behavior systematically, shifting the industry from a paradigm of ``trust without evidence'' to a paradigm of verifiable, cryptographically secure accountability \cite{b12,b17}.

\section{Future Release Roadmap and Scope}

The Evaluation Context Protocol described in this paper is at an early stage, and the reference implementation carries an experimental status label \cite{b14}. The protocol currently specifies a deliberately minimal surface: three JSON-RPC methods, one result envelope with three graded fields and one ungraded field, two transports, and three grader families \cite{b15}. That minimalism is a design choice rather than an omission, because a contract layer earns adoption only if it is small enough to be reimplemented independently. This section sets out the capabilities targeted for subsequent work, the boundaries the protocol does not intend to cross, and the empirical work required before ECP could be considered a settled standard. Table~\ref{tab:roadmap} groups these into indicative phases; we deliberately describe them as phases rather than as versioned releases, because the sequencing will depend on what early adopters find breaks first.

\subsection{The Evaluation Surface Is Expected to Change}
The most important thing to say about future work is that the evaluation surface itself is provisional. The three graded fields presented in Section~VI are not derived from a theory of what agent behavior consists of; they are the smallest set that let us express checks against the failure modes catalogued in Section~IV, and they were shaped by what existing frameworks could be made to emit through the adapters. A different starting point would plausibly have produced a different decomposition.

Several forces are likely to reshape it. Some evaluation needs do not fit any current field: the cost and latency of a step, the observation a tool returned, the identity of the subagent that acted, and the confidence an agent attached to its own answer are all things evaluators ask about and none are currently expressible. Some fields may prove to be doing too much at once; \texttt{evaluation\_context} in particular is currently a free-text channel serving as reasoning summary, policy evidence, and trace summary simultaneously, and it may need to be decomposed into distinct typed fields with separate disclosure rules. Others may prove unnecessary. The optional \texttt{logs} field, added because implementations wanted somewhere to put execution output, has no dedicated grader and may be folded elsewhere or removed.

The design principle we intend to hold to is not the specific number of fields but the separation underneath it: what the user sees, what the agent did, and what evidence the agent can safely expose about why, are distinct questions that deserve distinct assertions. How many fields best express that separation is an empirical question, and we expect to revise it. Any such change will require a versioning and deprecation discipline that the protocol does not yet have; the existing acceptance of a deprecated field alias as a compatibility shim is a precedent for how such transitions should be handled, but a general policy needs to be written down before the surface changes again.

\subsection{Expanding the Protocol Surface}
The present method set (\texttt{agent/initialize}, \texttt{agent/step}, \texttt{agent/reset}) models a single agent process advanced turn by turn \cite{b15}. Three extensions are prioritized.

First, \textit{capability negotiation} must become explicit. The \texttt{capabilities} object returned by \texttt{agent/initialize} is currently informational; a future revision will define a normative capability vocabulary so that a runtime can detect at handshake time whether an agent supports streaming, cancellation, or multi-agent attribution, and can degrade or fail loudly rather than silently mis-grading.

Second, the \texttt{tool\_calls} array records a tool name and its arguments but not the observation returned by the environment, the wall-clock duration, or the token cost of the invoking step \cite{b15}. Enriching each entry with an optional \texttt{result}, \texttt{latency\_ms}, and \texttt{cost} field would make trajectory efficiency, the dimension identified in Table~\ref{tab:dimensions}, a directly gradeable property rather than an inference drawn from external telemetry. This is the single change most likely to close the gap between ECP reports and the observability platforms compared in Table~\ref{tab:platforms}.

Third, and most consequentially, ECP currently has no first-class representation of \textit{delegation}. Multi-agent systems are precisely the setting in which architecture, rather than the backing foundation model, dominates observed performance \cite{b26,b27}. A future \texttt{agent/handoff} notification, together with a subagent identifier carried through the result envelope, would allow the runtime to reconstruct a delegation tree and to attribute a failed check to the specific subagent, handoff boundary, or shared-state mutation that produced it. Without this, ECP grades a multi-agent system as though it were a single opaque agent, which is the same aggregation error that trajectory-aware evaluation was introduced to avoid \cite{b1}.

A related item is the \texttt{paused} execution status already permitted by the specification. Its semantics are currently under-specified; formalizing it as a human-in-the-loop suspension point, with a defined resumption payload, would allow ECP to express the hybrid evaluation frameworks discussed in Section~III rather than only fully autonomous runs \cite{b17}.

\subsection{A Broader and Provider-Neutral Grader Ecosystem}
The reference runtime ships with \texttt{text\_match}, \texttt{llm\_judge}, and \texttt{tool\_usage} graders \cite{b15}. Four additions follow directly from the failure modes catalogued in Section~IV.

\textit{Statistical graders.} Section~IV-C argues that $pass@k$ systematically overstates operational reliability and that $pass^k$ is the appropriate production metric \cite{b16}. The runtime should therefore execute a scenario $k$ times natively, report $pass@k$ and $pass^k$ side by side with confidence intervals, and expose run-to-run variance as a first-class field. Reliability is a property of a distribution of runs, and a protocol that reports only a single run cannot express it \cite{b26}.

\textit{Policy and authority graders.} A declarative \texttt{policy\_check} grader would allow a manifest to express allow-lists, deny-lists, argument constraints, and ordering constraints over tool invocations, so that an agent which reaches the correct answer by touching a prohibited system fails deterministically rather than depending on a judge model to notice the violation \cite{b1}.

\textit{Deterministic execution graders.} A sandboxed \texttt{code\_exec} grader would restore the code-based verification that remains the ideal default for machine-checkable outcomes, reserving expensive model-based judging for the genuinely subjective residue \cite{b5}.

\textit{Calibrated judging.} The \texttt{llm\_judge} grader presently assumes a single commercial provider and reads one environment variable for credentials. Provider-neutral adapters are therefore required for ECP to be credibly vendor-neutral. Beyond portability, the governance thresholds discussed in Section~IV-D should become machine-checkable: a manifest should be able to declare the golden dataset a judge was calibrated against, the measured Cohen's $\kappa$, and the date of last recalibration, and the runtime should refuse to honor a judge whose calibration record is absent, below threshold, or stale \cite{b16}. A judge that cannot present its calibration record is, by that argument, a liability rather than an instrument.

\subsection{Provenance, Contamination Control, and Signed Evidence}
Section~IX-B anticipates a shift toward verifiable accountability, but the current release emits unsigned JSON and HTML artifacts that any party in the chain could edit \cite{b15}. Planned work therefore includes content-addressed hashing of the manifest, the agent revision, and the model identifiers used, together with detached cryptographic attestations over the emitted report. A third-party evaluator could then verify that a published result genuinely corresponds to the manifest and agent revision it claims, which is the minimum precondition for independent auditing \cite{b12}.

The protocol can also help enforce the developmental versus held-out separation that Section~IV-B identifies as necessary to prevent test contamination \cite{b17}. A sealed manifest mode, in which expected outputs are stored as commitments that the developing team cannot read but the runtime can verify, would render contamination detectable rather than merely discouraged by convention. Complementary work on redaction hooks for \texttt{evaluation\_context} is required so that audit evidence can be shared with external evaluators without leaking regulated personal data \cite{b14}.

\subsection{Interoperability and Multi-Language Conformance}
Three interoperability tracks are planned. Mapping the ECP result envelope onto the OpenTelemetry GenAI semantic conventions would allow a single instrumented run to populate both an ECP report and any OTEL-native observability backend, positioning ECP as complementary to, rather than competitive with, the platforms compared in Table~\ref{tab:platforms} \cite{b12,b25}. An MCP bridge would allow expected tool invocations to be derived automatically from the tool schemas an MCP server already publishes, removing a manual and error-prone step from manifest authoring \cite{b9,b13}. An early TypeScript SDK exists in the repository but is not yet at parity with the Python implementation \cite{b14}; bringing it to parity, and publishing a conformance profile that the existing harness can certify against, is required before the claim of language neutrality is more than architectural. A protocol is only vendor-neutral once at least two independent implementations pass the same conformance suite. The current export path to an external tracing platform should likewise be generalized, since a single-platform integration reproduces in miniature the coupling the protocol exists to avoid.

\subsection{Cost, Scale, and Domain Profiles}
Section~VIII-A notes that comprehensive evaluation can consume ten to fifteen percent of an organization's inference budget \cite{b16}. The planned mitigations are structural rather than incidental: parallel scenario sharding, content-addressed caching of judge verdicts so that unchanged scenarios are not re-judged, tiered execution in which inexpensive deterministic graders gate expensive model-based ones, and manifest-level budget ceilings that fail a run explicitly when exceeded rather than silently overspending.

Separately, the healthcare findings summarized in Section~VII-D suggest clear value in domain profiles: opinionated manifest templates that require the outcome and safety fields a given regulatory context expects, so that a clinical evaluation cannot quietly report only process and efficiency metrics \cite{b20,b24}.

\subsection{Scope Boundaries and Non-Goals}
Equally important is what ECP does not intend to become. ECP is not an observability backend, and does not store, index, or visualize traces at scale; that role belongs to the platforms surveyed in Section~V, which ECP aims to feed rather than replace \cite{b25,b28,b29}. ECP is not a benchmark or a task corpus, and takes no position on which tasks constitute a good evaluation. ECP is not an agent orchestration framework, and does not schedule, route, or supervise agents in production. Most importantly, ECP does not mandate the disclosure of raw chain-of-thought: the \texttt{evaluation\_context} field is defined as evaluator-safe structured evidence precisely so that providers can support trajectory-level auditing without exposing proprietary reasoning traces \cite{b10,b14}, and any future revision that erodes that boundary would remove the incentive for closed-model providers to participate at all.

\subsection{Threats to Validity and the Empirical Agenda}
This paper argues for ECP on architectural grounds and situates it against documented failure modes in the literature; it does not yet present a controlled evaluation of the protocol itself. That is the principal limitation of the present work. The planned empirical agenda has three parts. First, a fault injection study, in which known trajectory defects such as unauthorized tool use, fabricated intermediate values, and inefficient recovery loops are deliberately introduced into otherwise correct agents, and the detection rate of the full ECP surface is compared against outcome-only grading; the quantity of interest is the proportion of injected faults that yield a correct final answer yet still fail an ECP check. Second, a portability study running an identical manifest against functionally equivalent agents implemented in several frameworks, measuring verdict agreement in order to test the central portability claim. Third, an overhead study quantifying the wall-clock and monetary cost that ECP adds relative to native framework evaluation, since a contract layer that is correct but prohibitively expensive will not be adopted \cite{b16}. Until these results exist, ECP should be read as a proposed standard with a working reference implementation rather than as an empirically validated one.

\begin{table}[htbp]
\caption{Indicative Development Phases for the Evaluation Context Protocol}
\label{tab:roadmap}
\centering
\footnotesize
\begin{tabularx}{\columnwidth}{@{}l X@{}}
\toprule
\textbf{Phase} & \textbf{Representative Deliverables} \\
\midrule
Current &
Three JSON-RPC methods; \texttt{stdio} and Streamable HTTP transports; \texttt{text\_match}, \texttt{llm\_judge}, and \texttt{tool\_usage} graders; adapters for four agent frameworks; JSON and HTML reports; conformance harness; published JSON Schemas. \\
\addlinespace
Rigor &
Native repeated execution with $pass@k$ and $pass^k$ reporting; variance and flakiness metrics; enriched tool call entries carrying results, latency, and cost; a declarative policy and authority grader; sandboxed deterministic graders. \\
\addlinespace
Trust &
Provider-neutral judge adapters; machine-checkable judge calibration records; signed reports and content-addressed manifest provenance; sealed held-out manifests; redaction hooks for audit evidence. \\
\addlinespace
Ecosystem &
Delegation and handoff semantics with per-subagent attribution; a structured schema for evaluation context; formalized human-in-the-loop suspension; general observability export rather than a single platform; maturation of the TypeScript SDK. \\
\addlinespace
Stability &
A versioning and deprecation policy for surface changes; frozen normative specification; at least two independent implementations passing one conformance profile; regulated-domain manifest profiles; published empirical validation. \\
\bottomrule
\end{tabularx}
\end{table}

\section{Conclusion}

The evaluation of autonomous AI agents has decisively outgrown the static, prompt-response benchmarks designed for early foundation models. Because modern agents are highly dynamic systems that plan, execute external tools, manage persistent memory, and adapt to non-deterministic environments over extended operational horizons, their true efficacy and safety can only be assessed through comprehensive trajectory analysis. Continuing to evaluate agents solely based on their final textual output critically obscures systemic logic flaws, hallucinatory tool usage, and dangerous operational shortcuts.

Addressing this monumental challenge requires deep investment in specialized observability platforms capable of parsing massive, nested telemetry at scale. It also requires the adoption of stringent statistical methodologies, such as $pass^k$, to measure consistent reliability rather than best-case capability. Crucially, the field requires standardized integration and evaluation layers to overcome severe infrastructure fragmentation. 

While the Model Context Protocol (MCP) has provided a widely adopted language for agents to interface with external tools, this paper argues that an analogous contract is missing on the evaluation side, and proposes the Evaluation Context Protocol (ECP) as an early candidate for that role. By separating the evaluation contract from the underlying execution platform through a small JSON-RPC interface, and by exposing the agent's user-visible output, its tool calls, and evaluator-safe audit context as independently gradeable fields, ECP allows behavioral checks to be embedded directly in continuous integration pipelines and to be ported across frameworks.

We close by restating the limits of this contribution. ECP is at an early stage. Its evaluation surface reflects the failure modes we set out to catch rather than a settled theory of agent behavior, and we expect the fields, methods, and grader families to change as the protocol meets systems we have not anticipated; the framework adapters described in Section~VI demonstrate portability but also expose how unevenly existing frameworks surface the evidence the protocol asks for. The empirical work that would justify adoption, in particular a controlled comparison against outcome-only grading and a demonstration that independent implementations agree, remains ahead of us. What we offer here is a concrete, runnable proposal and an argument for why the agentic ecosystem needs a shared evaluation contract at all. As multi-agent systems scale into high-stakes domains, we believe that some standardized, trajectory-aware evaluation layer will become necessary; whether ECP in its present form is the right one is a question we hope this paper opens rather than closes.

\end{document}